\documentclass[spanish,english]{revtex4-1}
\usepackage[T1]{fontenc}
\usepackage[latin9]{inputenc}
\usepackage{verbatim}
\usepackage{amsmath}
\usepackage{amssymb}
\usepackage{graphicx}
\usepackage{esint}

\makeatletter
\renewcommand\[{\begin{equation}}
\renewcommand\]{\end{equation}} 

\makeatother

\usepackage{babel}
\addto\shorthandsspanish{\spanishdeactivate{~<>}}

\begin{document}
\global\long\def\abs#1{\left| #1 \right| }
\global\long\def\ket#1{\left| #1 \right\rangle }
\global\long\def\bra#1{\left\langle #1 \right| }
\global\long\def\half{\frac{1}{2}}
\global\long\def\partder#1#2{\frac{\partial#1}{\partial#2}}
\global\long\def\comm#1#2{\left[ #1 ,#2 \right] }
\global\long\def\vp{\vec{p}}
\global\long\def\vpp{\vec{p}\, ^{\prime}}
\global\long\def\dt#1{\delta^{(3)}(#1 )}
\global\long\def\Tr#1{\textrm{Tr}\left\{  #1 \right\}  }
\global\long\def\Real#1{\mathrm{Re}\left\{  #1 \right\}  }
\global\long\def\braket#1{\langle#1\rangle}
\foreignlanguage{spanish}{}\global\long\def\escp#1#2{\left\langle #1|#2\right\rangle }
\foreignlanguage{spanish}{}\global\long\def\elmma#1#2#3{\langle#1\mid#2\mid#3\rangle}

\title{Asymptotic properties of the Dirac quantum cellular automaton}

\author{A Pérez }

\affiliation{Departament de Física Teòrica and IFIC, \\
Universitat de València-CSIC, Dr. Moliner 50, 46100-Burjassot, Valencia
\\
Spain}
\begin{abstract}
We show that the Dirac quantum cellular automaton {[}Ann. Phys. 354
(2015) 244{]} shares many properties in common with the discrete-time
quantum walk. These similarities can be exploited to study the automaton
as a unitary process that takes place at regular time steps on a one-dimensional
lattice, in the spirit of general quantum cellular automata. In this
way, it becomes an alternative to the quantum walk, with a dispersion
relation that can be controlled by a parameter, which plays a similar
role to the coin angle in the quantum walk. The Dirac Hamiltonian
is recovered under a suitable limit. We provide two independent analytical
approximations to the long term probability distribution. It is shown
that, starting from localized conditions, the asymptotic value of
the entropy of entanglement between internal and motional degrees
of freedom overcomes the known limit that is approached by the quantum
walk for the same initial conditions, and are similar to the ones
achieved by highly localized states of the Dirac equation.
\end{abstract}
\maketitle

\section{Introduction}

The connection between physical processes on a lattice, and the corresponding
theories in the continuum, is intriguing and plagued with difficulties
and new features \cite{Muenster2007,Smit2002c,Montvay1994a}. Discretization
of quantum field theories that are defined on the continuum can be
regarded as a powerful calculation tool, a paradigmatic example being
QCD on a lattice \cite{Gupt2007}, that allows for non perturbative
calculations, after a suitable extrapolation is made to the limit
of vanishing lattice spacing. In the case of fermion fields, one encounters
problems like the ``fermion doubling'', which can be attacked in
different ways. This clearly shows that the discretization procedure
of quantum field theories is not uniquely defined, with different
approaches leading to the same limit in the continuum. In particular,
this is true for the Dirac equation, which describes the relativistic
motion of a spin 1/2 particle, and gives rise to interesting phenomena
as the \textit{Zitterbewegung} or the Klein paradox \cite{thaller}.

A recent paper \cite{Bisio2015} introduces a Dirac Quantum Cellular
Automaton (DQCA), that describes the relativistic dynamics of a spin
1/2 particle on a one-dimensional lattice based on some symmetry principles.
Quantum cellular automata have been studied by several authors (see,
for example \cite{PhysRevD.49.6920,Meyer1996a,Meyer1996,PhysRevE.55.5261,Meyer1998,Arrighi2011,Gross2012,Tosini2014}).
The model described in \cite{Bisio2015} can be regarded as a particular
case of the two component cellular automaton defined in \cite{Meyer1996a},
and works as a set of updating rules on discrete space-time coordinates,
where the time step and lattice spacing are to be identified with
the Planck time $\tau_{P}$ and Planck length $l_{P}$, respectively.
In the limit of large wavelengths (as compared to $l_{P}$) and small
masses $m\ll m_{P}$, with $m_{P}$ the Planck mass, the Hamiltonian
representing the DQCA approximates the Dirac Hamiltonian. The model
also accounts for the above mentioned \textit{Zitterbewegung} and
Klein paradox phenomena \cite{Bisio2013}. 

Also interesting is the fact that the evolution of the probability
distribution \cite{Bisio2015} resembles the one of a discrete time
Quantum Walk (QW). The QW is the quantum analogue of the classical
random walk. As in the case of random walks, QWs can appear either
under its discrete-time \cite{Aharonov93} or continuous-time \cite{PhysRevA.58.915}
form. Moreover, it has been shown that any quantum algorithm can be
recast under the form of a QW on a certain graph: QWs can be used
for universal quantum computation, this being provable for both the
continuous \cite{PhysRevLett.102.180501} and the discrete version
\cite{PhysRevA.81.042330}. Several experimental setups have been
already performed to implement the QW \cite{PhysRevLett.100.170506,PhysRevLett.103.090504,PhysRevA.67.042305,Karski2009,PhysRevLett.104.153602,PhysRevLett.104.050502,PhysRevLett.104.100503,Kitagawa2012,PhysRevLett.108.010502,Schreiber06042012,Crespi2013,Fukuhara2013,Xue2014,Cardano2014}
(for a comprehensive review, see \cite{Manouchehri:2013:PIQ:2566741}).

In addition to the probability distribution, one immediately finds
that the dispersion relation of the QW can be mapped into the one
corresponding to the DQCA. Last, but not least, both models reproduce
the Dirac equation in some limit, a property that has been established
by several authors in the case of the QW \cite{Strauch2006,Strauch2007,Kurzynski2008,Chandrashekar2013,Shikano2013,Arrighi2013,DiMolfetta2015}.
These similarities suggest that the two models may share other properties
that are worth studying. This is precisely the motivation of this
paper. We found some subtleties that will be discussed in detail,
once the long term evolution has been derived. Therefore, we establish
a link between a model motivated from a lattice field theory, on the
one side, and a process (the QW) that plays an important role in the
theory of quantum information.

This paper is organized as follows. In Sect. II we review the general
properties of the DQCA and the QW. The dispersion relations of both
models are discussed in Sect. III, and we show that the Dirac Hamiltonian
is obtained from a suitable limit of the DQCA unitary operator. The
similarities and differences of the probability distributions for
both models are analyzed in Sect. IV. In Sect. V, we derive two different
approximations to the long term probability distribution of the DQCA:
We first obtain a simple formula from the $r$-th moment of the position
operator at large time steps, which only describes the gross features
of the probability distribution, although we can extract the correct
analytical behavior of the standard deviation. We next obtain an approximate
result with the help of the stationary phase method, which turns out
to work very well, and correctly describes the details of the oscillations
in the probability. Sect. VI is devoted to the study of the entanglement
between the spatial and internal degrees of freedom, as quantified
by the entropy of entanglement. We will show that, for a localized
initial condition, this magnitude saturates the allowed maximum value
for a two-dimensional Hilbert space, at variance with the lower limiting
value which is approached by the QW for the same initial conditions.
We discuss the similarity of the obtained result with highly localized
initial states for the Dirac equation.

\section{General properties of the DQCA and the QW}

\subsection{QW}

The standard QW corresponds to the discrete (both in time and in space)
evolution of a one-dimensional quantum system (the walker) in a direction
which depends on an additional degree of freedom, the chirality, with
two possible states: ``left\textquotedblright{} $|L\rangle$ or ``right\textquotedblright{}
$|R\rangle$. The global Hilbert space of the system is the tensor
product $H_{s}\otimes H_{c}$. $H_{s}$ is the Hilbert space associated
to the motion on the line, and it is spanned by the basis $\{|x=nd\rangle:n\in\mathbb{Z}\}$,
where $d$ is the lattice spacing, usually taken as the unit length.
$H_{c}$ is the chirality (or coin) Hilbert space, defined as a two-dimensional
space that may correspond, for example, to a spin 1/2 particle, or
to a 2-level energy system. Let us call $T_{-}$ ($T_{+}$) the operators
in $H_{s}$ that move the walker one site to the left (right), and
$|L\rangle\langle L|$, $|R\rangle\langle R|$ the chirality projector
operators in $H_{c}$. We consider the unitary transformation 
\begin{equation}
U_{QW}=\left\{ T_{-}\otimes|L\rangle\langle L|+T_{+}\otimes|R\rangle\langle R|\right\} \circ\left\{ I\otimes C(\theta)\right\} ,\label{UQW}
\end{equation}
where $C(\theta)$ is the \textit{coin operator}, which acts only
on the coin space, and $I$ the identity operator in $H_{s}$. Any
$SU(2)$ matrix can be used but, for our purposes, it is sufficient
to parametrize 
\begin{equation}
C(\theta)=\left(\begin{array}{cc}
\cos\theta & -\sin\theta\\
\sin\theta & \cos\theta
\end{array}\right),\label{eq:coinoperator}
\end{equation}
with $\theta\in\left[0,\pi/2\right]$ a parameter defining the bias
of the coin toss. The effect of the unitary operator $U_{QW}$ on
the state of the system in one time step $\tau$ is $|\psi(t+\tau)\rangle=U_{QW}|\psi(t)\rangle$.
The state vector can be expressed as 
\begin{equation}
|\psi(t)\rangle=\sum\limits _{n=-\infty}^{\infty}\ket{nd}\otimes\left[a_{n}(t)\ket R+b_{n}(t)\ket L\right].
\end{equation}
From the above we obtain 
\begin{equation}
\ket{\psi(n,t)}\equiv\braket{nd|\psi(t)}=a_{n}(t)\ket R+b_{n}(t)\ket L\label{spinor}
\end{equation}
or, in vector notation $\ket{\psi(n,t)}=(a_{n}(t),b_{n}(t))^{T}.$
At any given time step, the probability distribution of the walker
can be calculated from
\begin{equation}
P(n,t)=\left\vert a_{n}(t)\right\vert ^{2}+\left\vert b_{n}(t)\right\vert ^{2}.\label{eq:probdistr}
\end{equation}

\subsection{DQCA}

As mentioned in the Introduction, the Dirac Quantum Cellular Automaton
is an extension of the Dirac field theory to the Planck and ultrarelativistic
scales \cite{Bisio2015,Bisio2013}. The model is defined by the repeated
action of a unitary operator $U_{DA}$ that acts on a spinor field
$\psi(x,t)$ with two internal degrees of freedom on a one-dimensional
lattice with spacing $l_{P}$ at time intervals $\tau_{P}$, $l_{P}$
and $\tau_{P}$ being the Planck length and Planck time, respectively.
In other words, $x=nl_{P}$, $t=k\tau_{P}$ with $n,k\in{\cal Z}.$
The state $\ket{\psi(t+\tau_{P})}$ of the system at time $t+\tau_{P}$
is related to the one at time $t$ by
\begin{equation}
\ket{\psi(t+\tau_{P})}=U_{DA}\ket{\psi(t)}.
\end{equation}
If we represent the two internal degrees of freedom by \{$|R\rangle$
,$|L\rangle$\}, as in the QW, then using similar steps we can define%
{} 
\begin{equation}
\ket{\psi(x,t)}=\braket{x|\psi(t)}=\psi_{R}(x,t)\ket R+\psi_{L}(x,t)\ket L,
\end{equation}
where $\psi_{R}(x,t)$ and $\psi_{L}(x,t)$ are the right and left
spinor components of $\ket{\psi(x,t)}$, respectively. In vector notation,
$\psi(x,t)=(\psi_{R}(x,t),\psi_{L}(x,t))^{T}.$

Starting from the hypothesis of unitarity, homogeneity of the interaction
topology, invariance under time reversal and parity, and minimal dimension
for a non-identical evolution, one arrives \cite{Bisio2015} to a
unitary operator that can be written as:

\begin{equation}
U_{DA}=\sqrt{1-\beta{}^{2}}\left\{ T_{-}\otimes|L\rangle\langle L|+T_{+}\otimes|R\rangle\langle R|\right\} -i\beta\sigma_{x},\label{UCA}
\end{equation}
with $\beta$ a real number. As already proven in \cite{Meyer1996a},
in one spatial dimension the conditions of homogeneity and locality
give rise to a \textit{no-go lemma} that prevents the existence of
nontrivial scalar quantum cellular automata. In other words, every
band $r$-diagonal unitary matrix $U$ which commutes with the 1-step
translation matrix $T_{+}$ is also a translation matrix $T_{+}^{k}$
for some $k\in\mathbb{Z}$, times a phase. One way to evade the lemma
is by combining two consecutive sites of the lattice on a cell, and
allow the cells to evolve and communicate (the so-called ``partitioning/alternating
evolution rule''). A more interesting possibility is the combination
of the two amplitudes of the cell on a field $\psi(x,t)$ with two
or more components. This allows to establish connections with field
theories (e.g., the Dirac equation). The simplest non-trivial case
being associated with a value $r=1$, and described by a two component
field. In this case, under the assumptions of locality, unitarity
and parity invariance it is shown in the above reference that one
arrives to the evolution rule 
\begin{equation}
\psi(t+1,x)=w_{-1}\psi(t,x-1)+w_{0}\psi(t,x)+w_{+1}\psi(t,x+1),\label{eq:Meyermap}
\end{equation}
where $w_{-1}$, $w_{0}$, and $w_{+1}$ are $2\times2$ matrices.
Any nontrivial solution to the above evolution rule turns out to be
unitarily equivalent to the choice 
\begin{equation}
w_{-1}=\cos\rho\left(\begin{array}{cc}
0 & i\sin\theta\\
0 & \cos\theta
\end{array}\right),\,\,\,w_{+1}=\cos\rho\left(\begin{array}{cc}
\cos\theta & 0\\
i\sin\theta & 0
\end{array}\right),\,\,\,w_{0}=\sin\rho\left(\begin{array}{cc}
\sin\theta & -i\cos\theta\\
-i\cos\theta & \sin\theta
\end{array}\right).\label{eq:Meyermatrices}
\end{equation}

It can be easily shown that the evolution defined by the DQCA Eq.
(\ref{UCA}) can be obtained from Eqs. (\ref{eq:Meyermap},\ref{eq:Meyermatrices})
making the choice $\theta=0$, $\beta=\sin\rho$. Thus, the DQCA can
be regarded as a particular case of the $r=1$ maps studied in \cite{Meyer1996a}.
As discussed in this reference, one can also relate this quantum cellular
automaton to the one dimensional version of Bialynicki-Birula's unitary
cellular automaton for the Dirac equation \cite{PhysRevD.49.6920}. 

Similarly to the QW, we can define the spatial probability distribution
as
\begin{equation}
P(n,t)=\left\vert \psi_{R}(nl_{P},t)\right\vert ^{2}+\left\vert \psi_{L}(nl_{P},t)\right\vert ^{2}.\label{eq:ProbDA}
\end{equation}

We want to establish a connection between both models. To this purpose,
we consider the original DQCA as a unitary operation taking place
on a lattice with arbitrary spacing $d$ at regular time steps $\tau$.
In other words, we replace 
\begin{equation}
l_{P}\rightarrow d,\,\,\,\tau_{P}\rightarrow\tau.\label{eq:aandtau}
\end{equation}

Notice that we depart from the original motivation of the DQCA as
a model to describe the relativistic behavior of spin 1/2 particle,
and consider the automaton as model that can potentially be realized
in the laboratory using similar setups as for the QW, and constitutes
an alternative to the latter. In what follows, we will investigate
the analogies and differences between both models.

We notice that the last term in Eq. (\ref{UCA}) only acts on the
internal degrees of freedom, and does not  include any displacement
on the lattice. As we show later, this introduces some characteristic
features on the evolution of the DQCA which are at variance with the
QW.

\section{Dispersion relation}

Most properties of the QW are better analyzed by switching to the
quasi-momentum space \cite{Nayak2007}. We introduce the basis of
states $\{\ket p,p\in[-\pi\hbar/d,\pi\hbar/d[\}$ defined by 
\begin{equation}
\ket p=\sqrt{\frac{d}{2\pi\hbar}}\sum\limits _{n=-\infty}^{\infty}e^{ipnd/\hbar}\ket{nd}.\label{eq:pstates}
\end{equation}
The unitary operators that govern both the QW and the DQCA become
diagonal in this basis. We represent these operators by $U_{QW}(p)$
and $U_{DA}(p)$, respectively. Furthermore, the internal indices
can be expressed in the \{$|R\rangle$, $|L\rangle$\} basis. With
these notations, we obtain 
\begin{equation}
U_{QW}(p)=\left(\begin{array}{cc}
e^{-ipd/\hbar} & 0\\
0 & e^{ipd/\hbar}
\end{array}\right)C(\theta)=\left(\begin{array}{cc}
e^{-ipd/\hbar}\cos\theta & -e^{-ipd/\hbar}\sin\theta\\
e^{ipd/\hbar}\sin\theta & e^{ipd/\hbar}\cos\theta
\end{array}\right),
\end{equation}
and
\begin{equation}
U_{DA}(p)=\left(\begin{array}{cc}
\sqrt{1-\beta{}^{2}}e^{-ipd/\hbar} & -i\beta\\
-i\beta & \sqrt{1-\beta{}^{2}}e^{ipd/\hbar}
\end{array}\right).\label{eq:UDAp}
\end{equation}
In both cases, the eigenvalues can be written as $\eta_{+}(p)\equiv e^{-i\lambda(p)},\eta_{-}(p)\equiv e^{i\lambda(p)}$,
where $\lambda(p)$ satisfies the dispersion relation 
\begin{equation}
\cos\lambda(p)=\cos\theta\cos(pd/\hbar),\label{dispQW}
\end{equation}
for the QW, and
\begin{equation}
\cos\lambda(p)=\sqrt{1-\beta{}^{2}}\cos(pd/\hbar)\label{dispDA}
\end{equation}
in the case of the DQCA. Therefore, both dispersion relations take
the same form, provided that we identify
\begin{equation}
\cos\theta\longleftrightarrow\sqrt{1-\beta{}^{2}}.\label{eq:equaldisper}
\end{equation}
Many features of the time evolution can be obtained directly from
the dispersion relation, such as the proportionality constant appearing
in the asymptotic behavior of the standard deviation \cite{ahlbrecht:042201},
or the design of desired asymptotic probability distributions in one
\cite{1367-2630-12-12-123022} or more dimensions \cite{Hinarejos2013a}.
Let us notice, however, that even if the correspondence defined Eq.
(\ref{eq:equaldisper}) is respected, so that both dispersion relations
become equivalent, the operators by $U_{QW}(p)$ and $U_{DA}(p)$
posses a different structure, which results in the differences that
are discussed in the next sections.

From the unitary operator Eq. (\ref{eq:UDAp}) one can extract the
corresponding Hamiltonian, similarly to \cite{Bisio2015}. We first
write $t=l\tau$, where $l\in\mathbb{N}$, and $\tau$ is the time
step. We then define the Hamiltonian $H(p)$ by 
\begin{equation}
U_{DA}^{l}(p)\equiv\exp[-\frac{i}{\hbar}l\tau H(p)].
\end{equation}
Following this definition, one finds
\begin{equation}
H(p)=\frac{\hbar\lambda(p)}{\tau\sin\lambda(p)}\left(\begin{array}{cc}
\sqrt{1-\beta{}^{2}}\sin(pd/\hbar) & \beta\\
\beta & -\sqrt{1-\beta{}^{2}}\sin(pd/\hbar)
\end{array}\right).
\end{equation}
Let us now rewrite 
\begin{equation}
\beta\equiv\frac{mdc}{\hbar},
\end{equation}
with $m$ a parameter with dimensions of mass. The Dirac Hamiltonian
is recovered in the limit $pd/\hbar\ll1$, $mdc/\hbar\ll1$. In this
limit, we have $\sin\lambda(p)\simeq\lambda(p)$, $\sin(pd/\hbar)\simeq pd/\hbar$
and $\sqrt{1-\beta{}^{2}}\simeq1$, so that
\begin{equation}
H(p)\simeq\frac{d}{c\tau}\left(\begin{array}{cc}
pc & mc^{2}\\
mc^{2} & -pc
\end{array}\right).
\end{equation}
By choosing the time step and the lattice spacing such that $\tau=d/c$,
one obtains the Dirac Hamiltonian, where $m$ can be identified with
the mass of the particle. The latter condition is a reminder of the
original model, where $d=l_{P}$ and $\tau=\tau_{P}$, obviously related
by $\tau_{P}=l_{P}/c$. In our proposal, these two parameters are
no longer related to the Planck scale, but the Dirac dynamics can
be recovered within the above restrictions. Considered as a process
that may approximately simulate a more general wave dynamics via discretization
on a lattice, one would need to set the length of the discretization,
which would determine the correspondence of the parameters of the
physical system under simulation with the ones of the model (in our
case, the value of $\beta$). Also, the value of the time step $\tau$
is obtained from the requirement of a given elapsed time $t$, and
the available number of time steps in the simulation.

\section{Probability distribution}

In spite of sharing a common dispersion relation, both models will
differ in several other aspects. For our analysis, we will fix some
of the parameters appearing in these models. The QW will be studied
using $\theta=\pi/4$ in (\ref{eq:coinoperator}), a choice that can
be mapped to the standard Hadamard coin. Moreover, we adopt the convention
$d=1$: In this way, we can label the site states as $\ket n$ in
both cases. To allow a comparison with the QW, as discussed in the
previous section, we will use the value $\beta=1/\sqrt{2}$ for the
following plots.

We first study the probability distribution, as defined in Eqs. (\ref{eq:probdistr})
and (\ref{eq:ProbDA}) . 
\begin{figure}
\includegraphics[width=6cm]{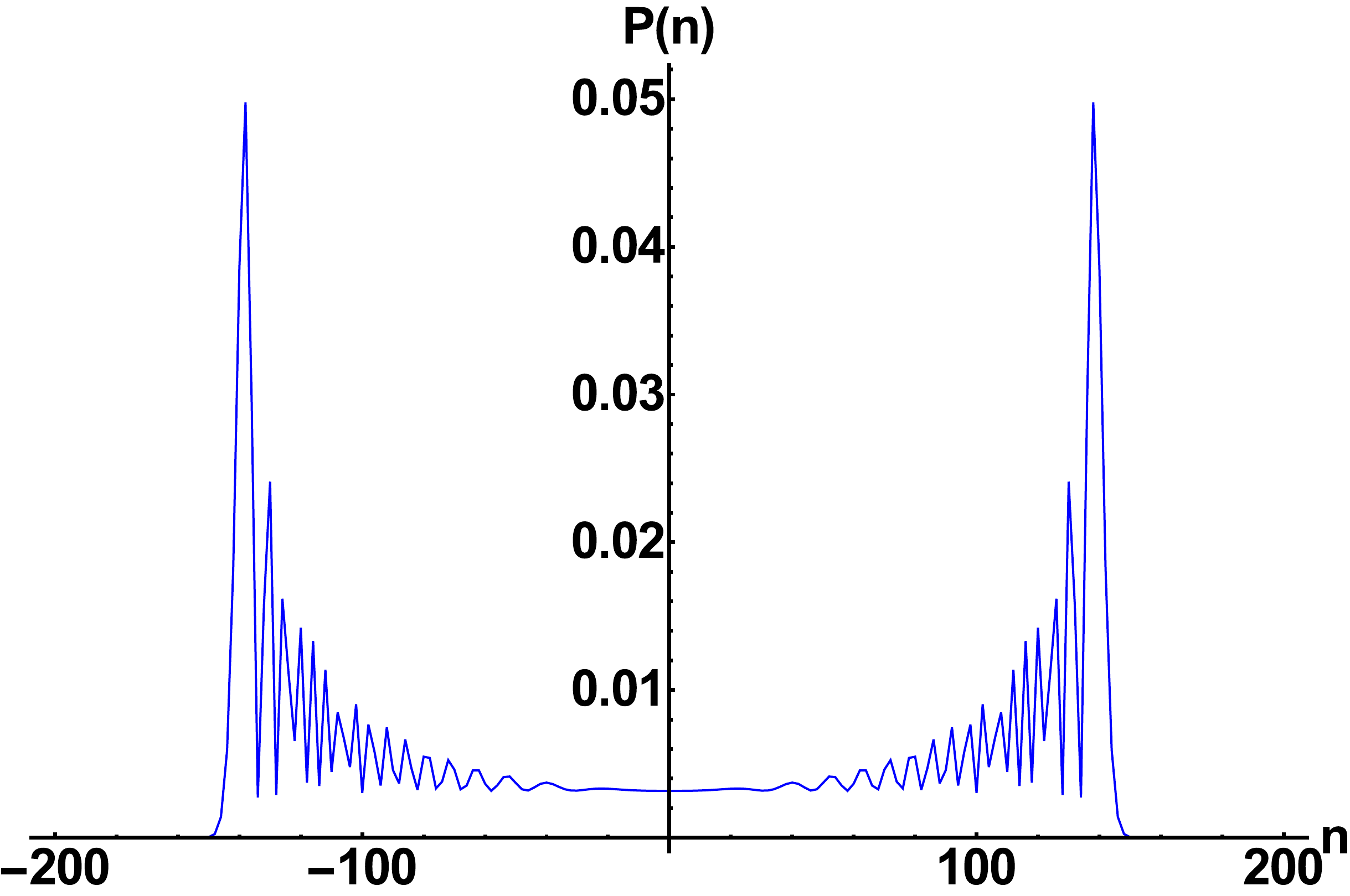}\includegraphics[width=6cm]{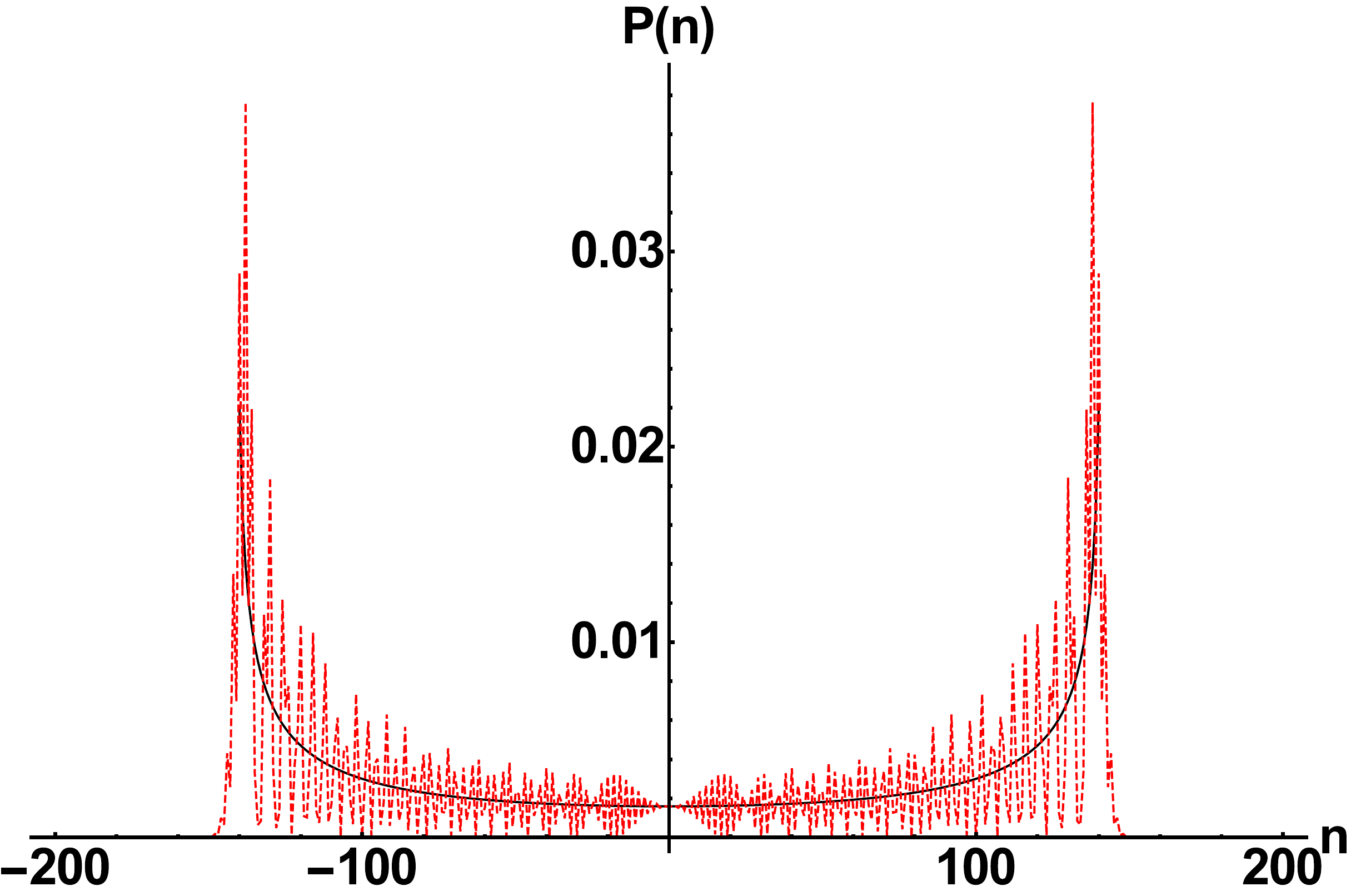}\includegraphics[width=6cm]{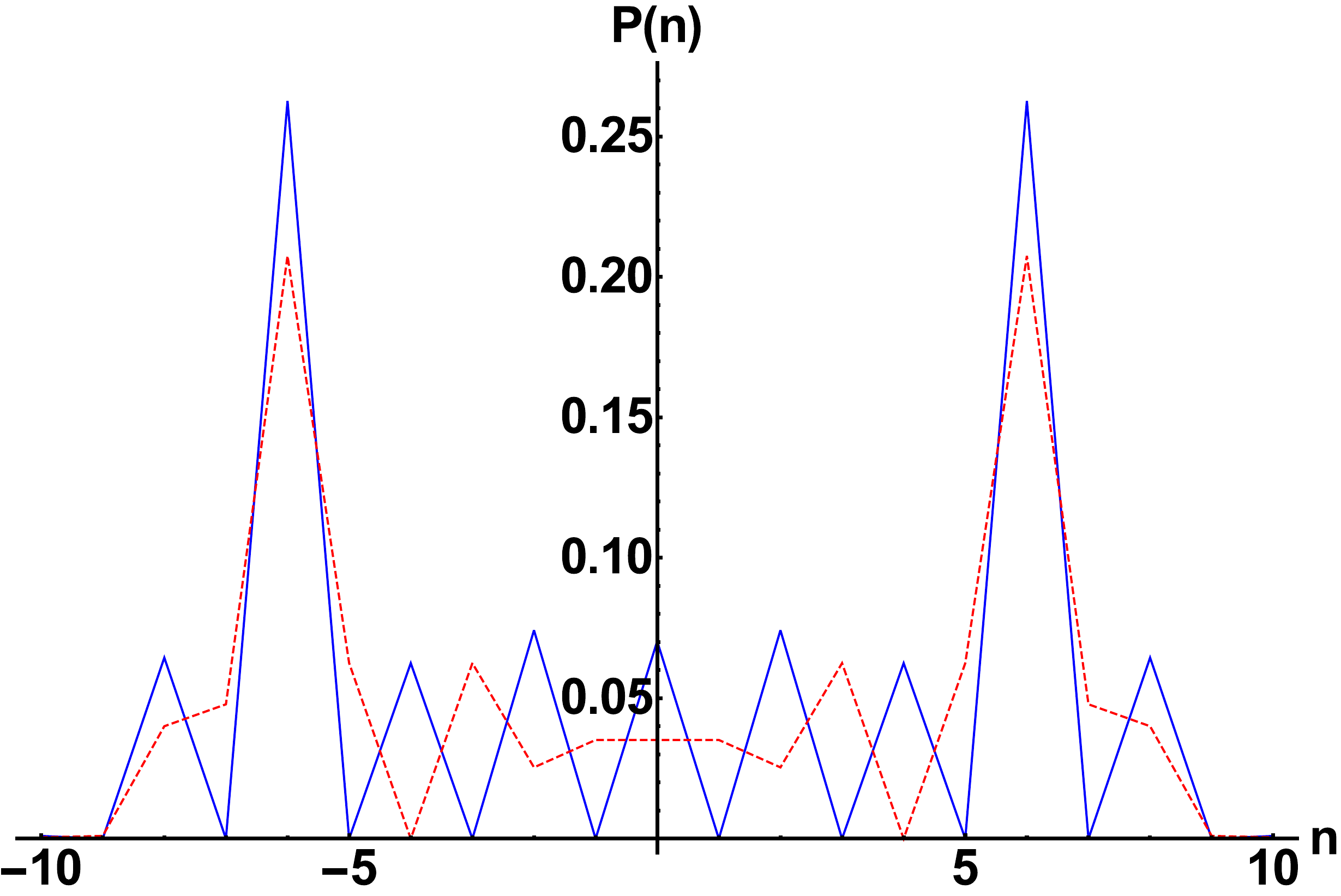}

\caption{(Color online) Left panel: Probability distribution after $t=200$
time steps for the QW, where only even sites are plotted. The initial
state is localized at the origin, see Eq. (\ref{eq:initialdelta}).
Middle panel: Distribution for the DQCA, with all sites showed (red
dashed line), compared with the long-term approximation, Eq. (\ref{eq:P(y)})
(black solid line). The right panel shows the differences for a smaller
($t=10)$ time step, where one clearly sees that the probability of
the QW vanishes at odd sites of the lattice.}

\label{Figprobdistr}
\end{figure}
Figure \ref{Figprobdistr} shows both probability distributions after
$t=200$ time steps, starting from the initial localized condition
\begin{equation}
\psi(n,0)=\frac{1}{\sqrt{2}}\left(\begin{array}{c}
1\\
i
\end{array}\right)\delta_{n,0}.\label{eq:initialdelta}
\end{equation}

One observes clear differences: The QW shows its characteristic peaks
and a flat distribution in the middle, whereas the DQCA features more
complicated structures. As it is well known, the probability distribution
of the QW vanishes at odd (even) sites of the lattice when $t$ is
even (odd). This is not true for the DQCA, as observed for $t=10$
on the same figure, the reason being that the last term in the unitary
transformation (\ref{UCA}) gives some probability to stay at the
same position, in contrast to the QW, where the particle is forced
to move left and right at each time step.

\begin{figure}
\includegraphics[width=10cm]{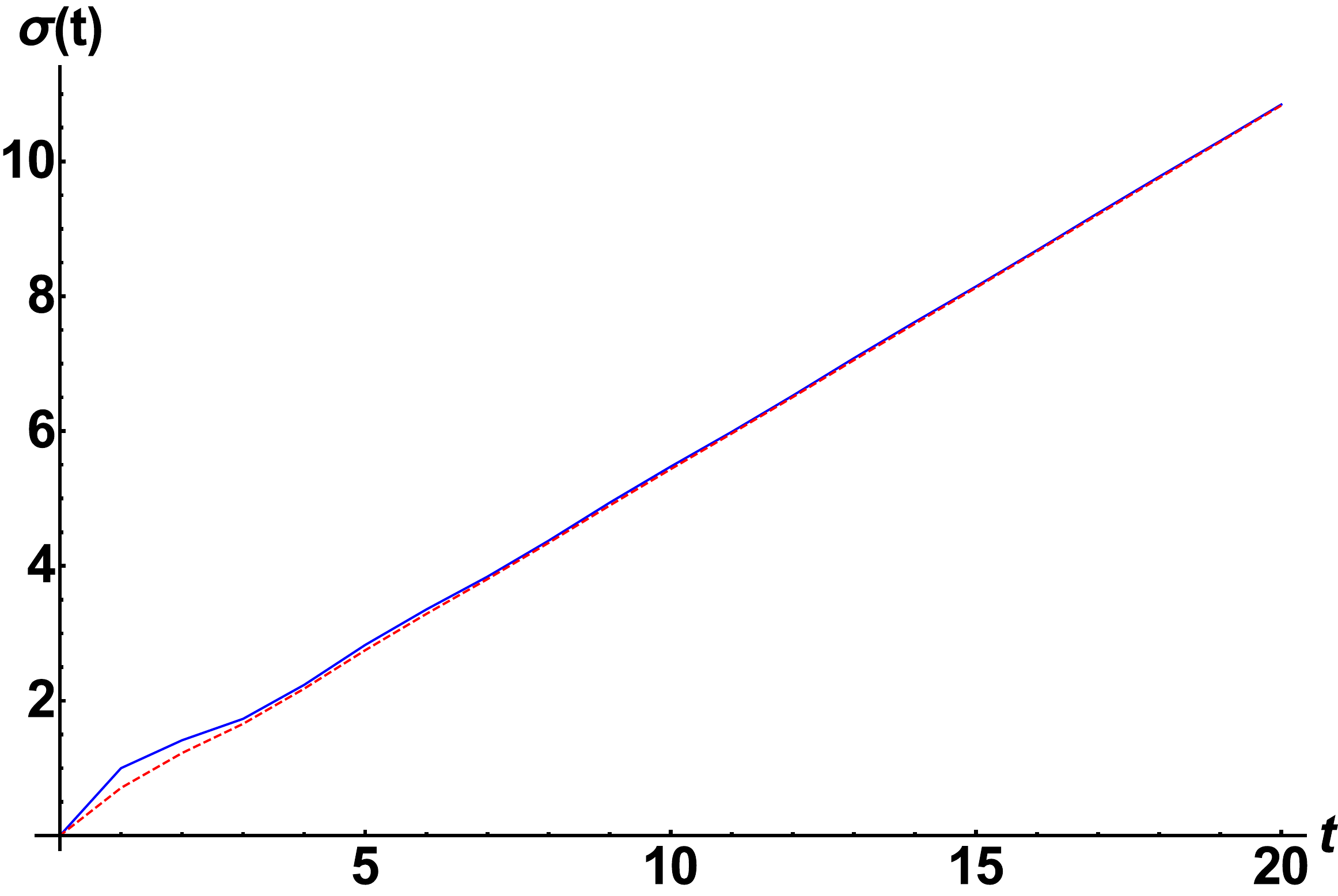}

\caption{(Color online) Standard deviation $\sigma(t)$ as a function of the
time step $t$ for the QW (blue solid line), and for the DQCA (red
dashed line). The initial state is localized at the origin, as in
the previous figure.}

\label{fig:sigma}
\end{figure}

Apart from these differences, the figure indicates that both probabilities
spread equally with time, at least for large time steps. In fact,
this is what Fig. (\ref{fig:sigma}) shows, where we plot the standard
deviation $\sigma(t)$ as a function of $t$: After a few time steps,
we obtain the characteristic ballistic $\sigma(t)\propto t$ spreading
of the QW in both cases.

\section{Asymptotic properties}

From the previous section it becomes apparent that, like in the QW
case, we can expect well defined properties for the DQCA at large
time steps. In fact, there are several methods to analytically derive
the long term behavior of the probability distribution. In what follows,
we will consider a localized initial state, as in the previous section.
Such narrow states in position space would pose problems in the original
formulation of the DQCA, i.e. when the model is intended to describe
the relativistic dynamics of a spin 1/2 particle, if one considers
a wave packet width smaller than the Compton wavelength of the particle
\cite{Bracken2005}. However, we have recast the DQCA as a discrete
time quantum process on an ordinary lattice, similar to the QW. In
this case, starting from a localized state (as compared to the lattice
spacing) can be realized in physical implementations. In fact, this
the most commonly studied situation, both theoretically and experimentally.
Therefore, in order to allow for a direct comparison, we consider
the localized state as our initial state. Also interesting is the
study of an initial Gaussian wave packet. As discussed above, this
becomes a necessity for the original DQCA motivation. The study of
Gaussian wave packets for the DQCA has been done in \cite{Bisio2015}.

\subsection{Weak limit }

We first make use of the method developed in \cite{PhysRevE.69.026119}
(see also \cite{Shikano2013}) to obtain the convergence of the $r$-th
moment 
\begin{equation}
E(x^{r},t)\equiv\braket{\psi(t)\mid x^{r}\mid\psi(t)},
\end{equation}
with $x$ the position operator (a different approach, based on combinatorial
methods, was used for the QW in \cite{springerlink:10.1023/A:1023413713008}).
Inserting the resolution of the identity in the basis of $\{\ket p\}$
states (\ref{eq:pstates}) one obtains
\begin{equation}
E(x^{r},t)=\int_{-\pi}^{\pi}dp\braket{\psi(p,t)\mid(i\frac{d}{dp})^{r}\mid\psi(p,t)},\label{eq:Exrt}
\end{equation}
where $\ket{\psi(p,t)}=\braket{p\mid\psi(t)}$ is a two-component
spinor in quasi-momentum space. Using the unitary operator (\ref{eq:UDAp})
in this basis, we can write
\begin{equation}
\ket{\psi(p,t)}=U_{DA}^{t}(p)\ket{\psi(p,0)}.\label{eq:iterpsip}
\end{equation}
The $t$-th power of $U_{DA}(p)$ is obtained from the spectral theorem:
\begin{equation}
U_{DA}^{t}(p)=\sum_{s=\pm1}e^{-is\lambda(p)t}\ket{\phi_{s}(p)}\bra{\phi_{s}(p)}.\label{eq:UDApt}
\end{equation}
In the latter equation, $\lambda(p)$ is obtained from the dispersion
equation (\ref{dispDA}), and $\ket{\phi_{s}(p)},s=\pm1$ are the
two normalized eigenvectors of $U_{DA}(p)$, given by \cite{Bisio2015}:
\begin{equation}
\ket{\phi_{s}(p)}=\frac{1}{\sqrt{2}}\left(\begin{array}{c}
\sqrt{1+sv(p)}\\
s\sqrt{1-sv(p)}
\end{array}\right),\label{eq:eigenp}
\end{equation}
where $v(p)=\frac{d\lambda}{dp}=\sqrt{1-\beta{}^{2}}\sin(p)/\sin\lambda(p)$
is the group velocity that arises from (\ref{dispDA}). Using Eqs.
(\ref{eq:iterpsip}) and (\ref{eq:UDApt}), one arrives to the following
relation 
\begin{equation}
\braket{\psi(p,t)\mid(i\frac{d}{dp})^{r}\mid\psi(p,t)}=(t)_{r}\sum_{s=\pm1}(\frac{i\eta'_{s}(p)}{\eta_{s}(p)})^{r}\mid\braket{\phi_{s}(p)\mid\psi(p,0)}\mid^{2}+{\cal O}(t^{r-1}).
\end{equation}
In this equation $\eta'_{s}(p)$ indicates the derivative with respect
to $p$, and $(t)_{r}\equiv t(t-1)\cdots(t-r+1)$. After dividing
by $t^{r}$ and taking the limit $t\longrightarrow\infty$, one obtains
\begin{equation}
\lim_{t\longrightarrow\infty}E(x^{r}/t^{r},t)=\sum_{s=\pm1}\int_{-\pi}^{\pi}dp\,\,(\frac{i\eta'_{s}(p)}{\eta_{s}(p)})^{r}\mid\braket{\phi_{s}(p)\mid\psi(p,0)}\mid^{2}.\label{eq:Exrwithp}
\end{equation}
Let us work out the above expression for the initial localized state
Eq. (\ref{eq:initialdelta}), for which $\ket{\psi(p,0)}=\frac{1}{2\sqrt{\pi}}\left(\begin{array}{c}
1\\
i
\end{array}\right).$ In this case, one finds $\mid\braket{\phi_{s}(p)\mid\psi(p,0)}\mid^{2}=1/4\pi$,
both for $s=1$ and $s=-1$. On the other hand, we can write $\frac{i\eta'_{s}(p)}{\eta_{s}(p)}=sv(p)$
, so that Eq. (\ref{eq:Exrwithp}) becomes 
\begin{equation}
\lim_{t\longrightarrow\infty}E(x^{r}/t^{r},t)=\frac{1}{4\pi}\int_{-\pi}^{\pi}dp\,\,[v^{r}(p)+(-v(p))^{r}].\label{eq:Exrwithp-1}
\end{equation}
We next change the integration variable in the first term of the latter
expression by inverting the function $y=v(p)$. Similarly, we perform
the transformation $y=-v(p)$ on the second term. After some algebra,
we arrive to the final expression 
\begin{equation}
\lim_{t\longrightarrow\infty}E(x^{r}/t^{r},t)=\frac{1}{\pi}\int_{-\sqrt{1-\beta^{2}}}^{\sqrt{1-\beta^{2}}}dy\,\,p'(y)y^{r},\label{eq:Exrwithp-1-1}
\end{equation}
with the notation $p'(y)\equiv\frac{|\beta|}{(1-y^{2})\sqrt{1-\beta^{2}-y^{2}}}$.
Eq. (\ref{eq:Exrwithp-1-1}) implies that the variable $x/t$ is distributed
across the interval $[-\sqrt{1-\beta^{2}},\sqrt{1-\beta^{2}}]$ with
a probability distribution given by 
\begin{equation}
P(y)=\frac{1}{\pi}p'(y)\equiv\frac{|\beta|}{\pi(1-y^{2})\sqrt{1-\beta^{2}-y^{2}}}.\label{eq:P(y)}
\end{equation}
As mentioned above, the same method has been applied to derive the
asymptotic probability distribution of the QW. We quote this result
for comparison. If one starts from a localized state 
\begin{equation}
\psi(n,0)=\left(\begin{array}{c}
a\\
b
\end{array}\right)\delta_{n,0},\label{eq:initialdeltageneral}
\end{equation}
with $|a|^{2}+|b|^{2}=1$, and a coin operator as defined in Eq. (\ref{eq:coinoperator}),
the corresponding distribution $P(y)$ can be written as \cite{springerlink:10.1023/A:1023413713008,Shikano2013}
\begin{equation}
P(y)=\frac{|\sin\theta|}{\pi(1-y^{2})\sqrt{\cos^{2}\theta-y^{2}}}[1-(|b|^{2}-|a|^{2}-\frac{\sin2\theta Re(ab^{*})}{\cos^{2}\theta})y],\label{eq:P(y)QW}
\end{equation}

valid for $|y|<|\cos\theta|$. For the particular case Eq. (\ref{eq:initialdelta}),
the above formula simplifies to
\begin{equation}
P(y)=\frac{|\sin\theta|}{\pi(1-y^{2})\sqrt{\cos^{2}\theta-y^{2}}},\label{eq:P(y)QW-1}
\end{equation}

which coincides with Eq. (\ref{eq:P(y)}), provided the identification
Eq. (\ref{eq:equaldisper}) is made.

As shown in Fig. \ref{Figprobdistr}, our result Eq. (\ref{eq:P(y)})
provides a simple approximation to the actual probability distribution
of the DQCA, although it does not reproduce the oscillations seen
on the true evolution. However, it can be used to obtain the standard
deviation at large time steps, as this magnitude does not depend on
the details of the distribution. By taking $r=2$ in Eq. (\ref{eq:Exrwithp-1-1})
one obtains
\begin{equation}
\sigma(t)=t\sqrt{1-|\beta|},
\end{equation}
which accounts for the ballistic spreading observed in Sect. IV. Indeed,
the previous result was expected from the similarity of the dispersion
relations of the QW (see \cite{ahlbrecht:042201}) and the DQCA, and
confirmed by the equivalence of Eqs. (\ref{eq:P(y)}) and (\ref{eq:P(y)QW-1}).

\subsection{Stationary phase method}

A better approximation to the long-time asymptotic distribution can
be obtained following the stationary phase method, as used for the
QW in \cite{Nayak2007}. Let us consider a localized initial condition,
such that 
\begin{equation}
\ket{\psi(p,0)}=\frac{1}{\sqrt{2\pi}}\left(\begin{array}{c}
a\\
b
\end{array}\right),\label{eq:initialcodab}
\end{equation}
and $|a|^{2}+|b|^{2}=1$. Making use of Eqs. (\ref{eq:iterpsip}),
(\ref{eq:UDApt}) and (\ref{eq:eigenp}), we arrive to $\ket{\psi(p,t)}=(\psi_{R}(p,t),\psi_{L}(p,t))^{T},$
where 
\begin{equation}
\psi_{R}(p,t)=\frac{1}{\sqrt{2\pi}}[a\cos\lambda(p)t-iav(p)\sin\lambda(p)t-ib\sqrt{1-v^{2}(p)}\sin\lambda(p)t]\label{eq:PsiRp}
\end{equation}
\begin{equation}
\psi_{L}(p,t)=\frac{1}{\sqrt{2\pi}}[b\cos\lambda(p)t+ibv(p)\sin\lambda(p)t-ia\sqrt{1-v^{2}(p)}\sin\lambda(p)t].\label{eq:PsiLp}
\end{equation}
The corresponding spinor in position space is obtained from
\begin{equation}
\psi_{R,L}(n,t)=\int_{-\pi}^{\pi}\frac{dp}{\sqrt{2\pi}}e^{ipn}\psi_{R,L}(p,t),\label{eq:Fouriertransform}
\end{equation}
where $n\in\mathbb{Z}.$ Let us introduce the notation $\alpha=n/t$,
and the functions 
\begin{equation}
I_{i}(\alpha,t)=\int_{-\pi}^{\pi}\frac{dp}{2\pi}e^{it(\lambda(p)+\alpha p)}g_{i}(p),\label{eq:defIialphat}
\end{equation}
with $g_{1}(p)=1$, $g_{2}(p)=v(p)$, and $g_{3}(p)=\sqrt{1-v^{2}(p)}$.
Then, the above result can be written as 
\begin{equation}
\psi_{R}(n,t)=a\Real{I_{1}(\alpha,t)}-a\Real{I_{2}(\alpha,t)}-ibIm\{I_{3}(\alpha,t)\}\label{eq:psiRnt}
\end{equation}
\begin{equation}
\psi_{L}(n,t)=b\Real{I_{1}(\alpha,t)}+b\Real{I_{2}(\alpha,t)}-iaIm\{I_{3}(\alpha,t)\}.\label{eq:psiLnt}
\end{equation}
Our goal is to obtain an approximation to the integrals $I_{i}(\alpha,t)$,
with $g_{i}(p)$ a smooth function. As seen from the definition, Eq.
(\ref{eq:defIialphat}), the integrand contains an oscillatory exponential,
specially for large values of $t$. For this kind of integrals, we
can make use of the stationary phase method \cite{Bleistein1975}.
Let us consider the phase $\Phi(p,\alpha)=\lambda(p)+\alpha p$ appearing
in these integrals, for a given value of $\alpha$, as a function
of $p$. The basic idea behind the method is to minimize these oscillations
by expanding the phase $\Phi(p,\alpha)$ around some convenient point
$p(\alpha)$: 
\begin{equation}
\Phi(p,\alpha)\simeq\Phi(p(\alpha),\alpha)+(p-p(\alpha))\left.\partder{\Phi(p,\alpha)}p\right|_{p(\alpha)}+\frac{1}{2}(p-p(\alpha))^{2}\left.\partder{^{2}\Phi(p,\alpha)}{p^{2}}\right|_{p(\alpha)}.\label{eq:expansionphase}
\end{equation}
In the latter equation, it becomes clear that the second term is responsible
for strong oscillations, provided that the derivative $\partder{\Phi(p,\alpha)}p$
is large. The idea is to minimize these strong oscillations by choosing
$p(\alpha)$ such that $\partder{\Phi(p,\alpha)}p$ vanishes. Therefore,
one needs to look for the roots $p_{i}(\alpha),i=1,2\dots$ of the
equation
\begin{equation}
\partder{\Phi(p,\alpha)}p=v(p)+\alpha=0.\label{eq:conditstatph}
\end{equation}
 Let us first assume $\alpha>0$. After careful inspection, we obtain
the roots $p_{1}(\alpha)=-p_{s}$, and $p_{2}(\alpha)=\pi+p_{s}$,
where $p_{s}\equiv\arccos\sqrt{(1-\beta^{2}-\alpha^{2})/[(1-\alpha^{2})(1-\beta^{2})]}$.
For the first solution, one needs to replace $\lambda(p)\longrightarrow\lambda(p_{s})\equiv\lambda_{s},\,\,\,\lambda^{''}(p)\longrightarrow\lambda^{''}(p_{s})=\sqrt{1-\beta^{2}-\alpha^{2}}(1-\alpha^{2})/|\beta|\ge0$,
while for the second solution we have to use $\lambda(p)\longrightarrow\pi+\lambda_{s},\,\,\,\lambda^{''}(p)\longrightarrow-\lambda^{''}(p_{s})$.
For $\alpha<0$, one needs to change $p_{s}\longrightarrow-p_{s}$.

As we show below, to our purposes it will suffice to concentrate on
$I_{1}(\alpha,t)$. After substitution of the above results, we obtain
the following approximation
\begin{equation}
I_{1}(\alpha,t)\simeq\frac{1}{\sqrt{2\pi t\lambda^{''}(p_{s})}}[e^{it\phi(\alpha)+i\pi/4}+e^{it(|\alpha|\pi-\phi(\alpha)+\pi)-i\pi/4}],\label{eq:I1approx}
\end{equation}
 where $\phi(\alpha)=\lambda_{s}-|\alpha|p_{s}$, the above expression
being valid for $|\alpha|\le\sqrt{1-\beta^{2}}$. To obtain $I_{2}(\alpha,t)$
we make use of condition (\ref{eq:conditstatph}). It then follows
$I_{2}(\alpha,t)\simeq-\alpha I_{1}(\alpha,t)$. Following a similar
argument, we arrive to $I_{3}(\alpha,t)\simeq\sqrt{1-\alpha^{2}}I_{1}(\alpha,t)$.
After taking the real part in Eq. (\ref{eq:I1approx}), and expanding
the two resulting cosinus functions, one can easily show that $Re\{I_{1}(\alpha,t)\}$
vanishes whenever $t+n$ is an odd integer number (remember the definition
$\alpha=n/t$, where $n\in\mathbb{Z}$). In practice, this means that
$Re\{I_{1}(\alpha,t)\}$ is zero at odd (even) lattice sites when
the time step $t$ is even (odd). The opposite result is found for
$Im\{I_{1}(\alpha,t)\}$: it becomes zero at odd (even) lattice sites
when the time step $t$ is odd (even). It can be checked that this
properties are indeed obeyed by the exact function. In other words,
the different terms in Eqs. (\ref{eq:psiRnt},\ref{eq:psiLnt}) ``alternate''
their contribution, for a given time step, as a function of $n$,
thus obtaining a dynamics in the probability distribution that differs
from the QW, as already discussed in Sect. IV.

One might wonder how the above expressions are modified if we want
to use a different system of units such that $d$ takes an arbitrary
value, as defined in Sect. IB. We will not repeat the procedure, and
just give the final answer, since the above calculations still hold,
if one introduces $q\equiv pd$ as the integration variable in (\ref{eq:Fouriertransform}).
In this way, one obtains $\psi_{R,L}(nd,t)$ as the left hand side
in Eqs. (\ref{eq:psiRnt},\ref{eq:psiLnt}). With this modification,
the rest of the above results remain unchanged (with $p$ replaced
by the new variable $q$). 

In order to test the accuracy of the above approximations, we will
analyze the function $I_{1}(\frac{x}{dt},t)$, which we extend to
arbitrary values of $x$, so that $x=nd$ with $n\in\mathbb{Z}$ correspond
to the lattice sites. A similar argument would apply to the extension
of the time step $t$ to arbitrary times, by changing $t\longrightarrow t/\tau$
in the function $I_{1}(\frac{x}{dt},t)$. Notice that changing the
value of $d$ to $d'$ corresponds to looking for a new value $x'$
in this function, such that $x'/d'=x/d$. Therefore, it is sufficient
to consider $d=1$ in what follows. Also, within the stationary phase
method the functions $I_{i}(\alpha,t)$ for $i=2,3$ are related to
$I_{1}(\alpha,t)$ in a simple way, so that it suffices for us to
consider the latter function.

\begin{figure}
\includegraphics[width=9cm]{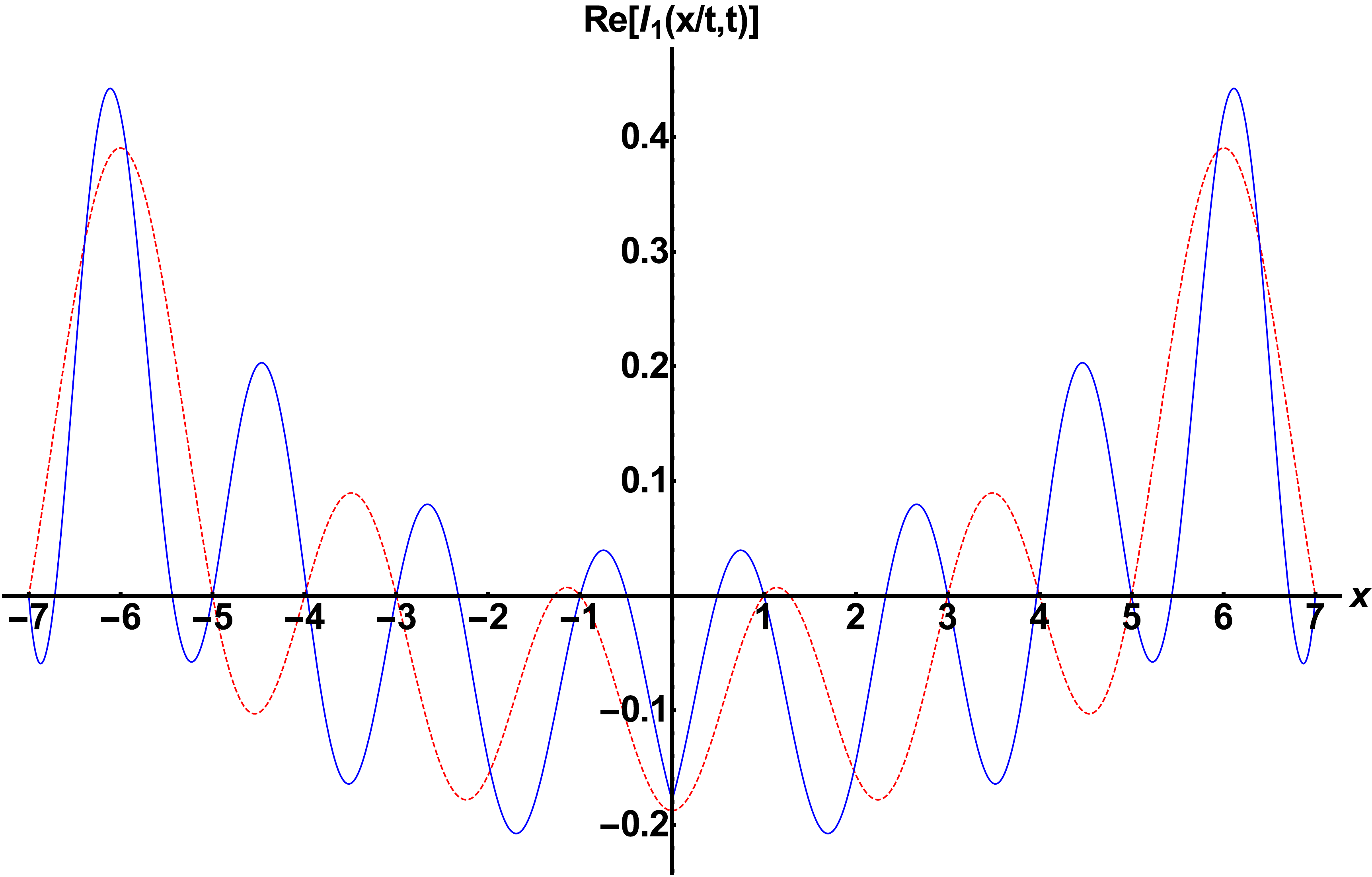}\includegraphics[width=9cm]{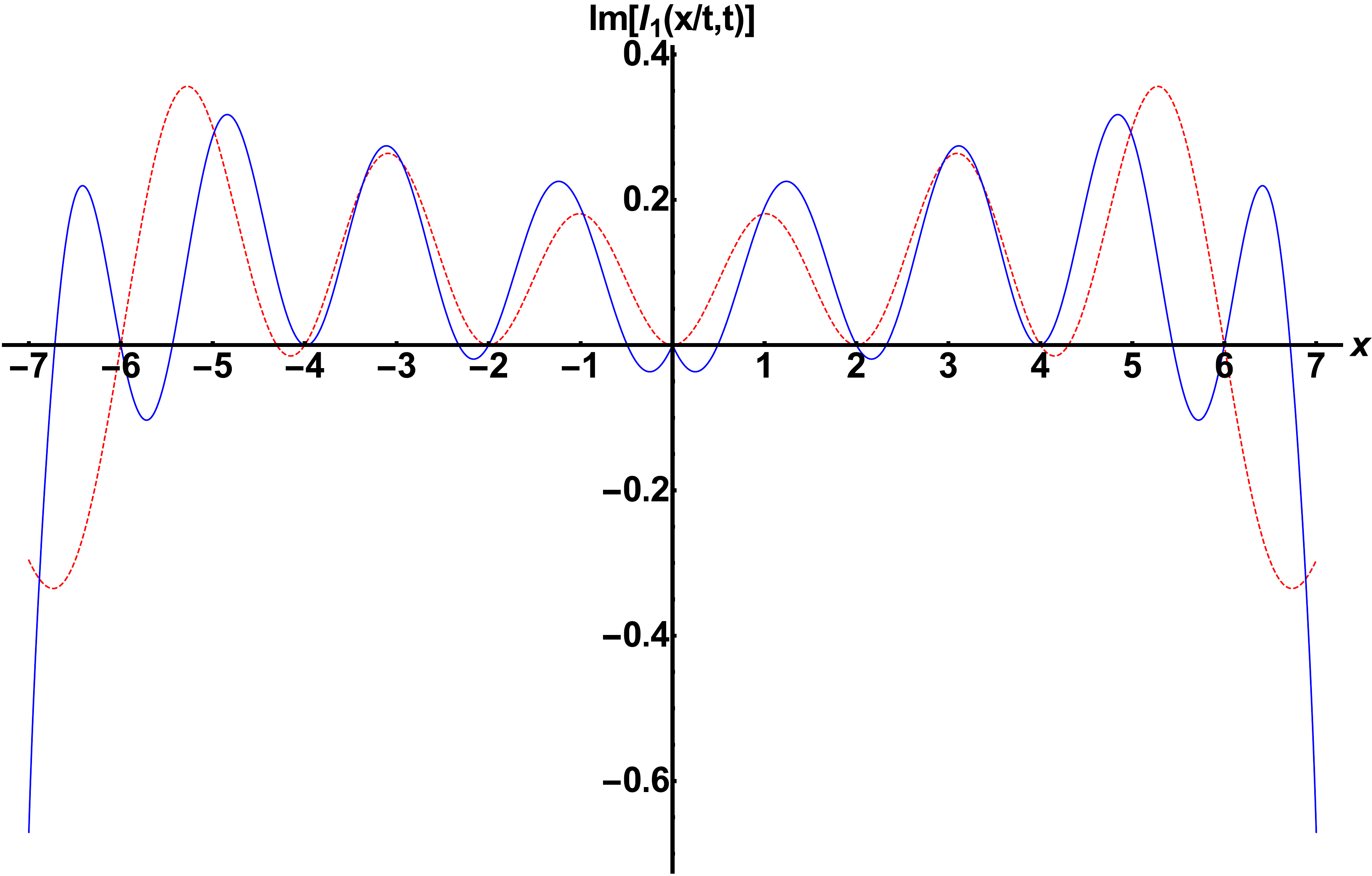}

\caption{(color online) Real part (left) and imaginary part (right) of the
function $I_{1}(\frac{x}{t},t)$ for $t=10$ time steps, as a function
of $x$. We adopted the value $\beta=1/\sqrt{2}$ . The dashed-red
curve was obtained from a numerical integration of Eq. (\ref{eq:defIialphat}),
while the blue solid curve corresponds to the obtained approximation
Eq. (\ref{eq:I1approx}). }

\label{figI1xt}
\end{figure}

\begin{figure}
\includegraphics[width=9cm]{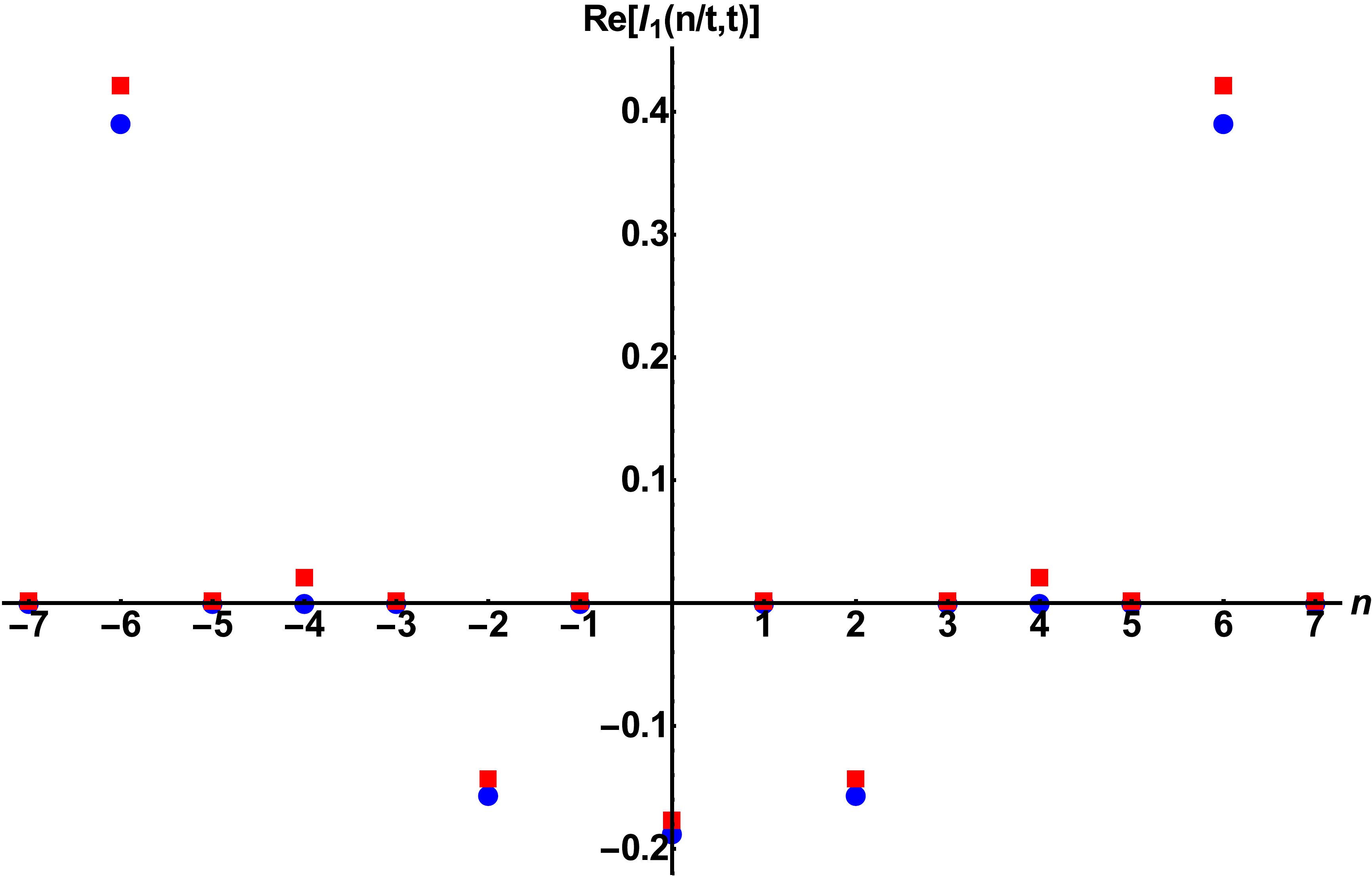}\includegraphics[width=9cm]{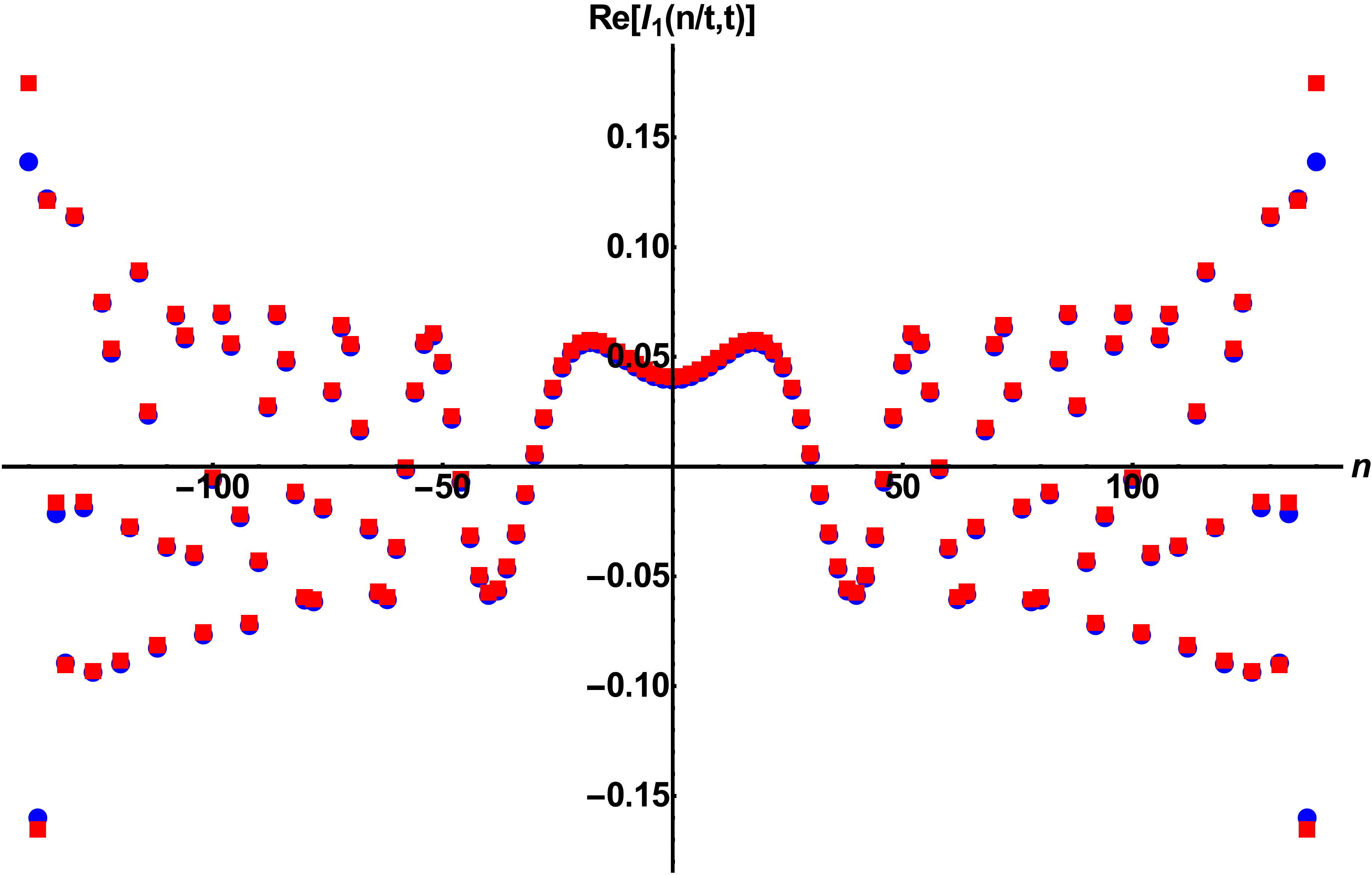}

\caption{(Color online) Real part of the function $I_{1}(\frac{x}{t},t)$,
where only lattice sites ($x=n$) are plotted. Blue dots are obtained
from numerical integration, whereas red squares correspond to the
approximate formula. The same value $\beta=1/\sqrt{2}$ was used.
The left panel corresponds to $t=10$, while the right panel is for
$t=200$. In the latter plot, only even sites are represented for
a better visualization, since the function vanishes at odd sites.}

\label{figtablasI1}
\end{figure}

Fig. \ref{figI1xt} shows the real and imaginary parts of the function
$I_{1}(\frac{x}{t},t)$ for $t=10$ time steps, as obtained from direct
numerical integration of Eq. (\ref{eq:defIialphat}), compared to
the obtained approximation Eq. (\ref{eq:I1approx}). As can be seen
from the plots, both curves show only an overall resemblance at small
number of time steps. However, we show them in order to better appreciate
the above mentioned parity properties, i.e. in this case the real
part vanishes at odd sites, whereas the imaginary part does at even
sites. The agreement between both functions improves if ones restricts
to physical sites of the lattice (i.e., for points $x=nd$ such that
$n\in\mathbb{Z}$). This can be appreciated from Fig. \ref{figtablasI1},
where the same results as in Fig. \ref{figI1xt} are represented,
restricted to lattice sites. On the same figure, one can see that
the approximate formula for $I_{1}(\frac{x}{t},t)$ works better for
larger time steps, as expected from the stationary phase method, although
it deviates from the exact value as $|\alpha|$ approaches the maximum
$\sqrt{1-\beta^{2}}$. A similar degree of agreement can be found
if one extends $I_{1}(\frac{x}{t},t)$ to arbitrary (i.e., non integer)
values of $t$.

\begin{figure}
\includegraphics[width=10cm]{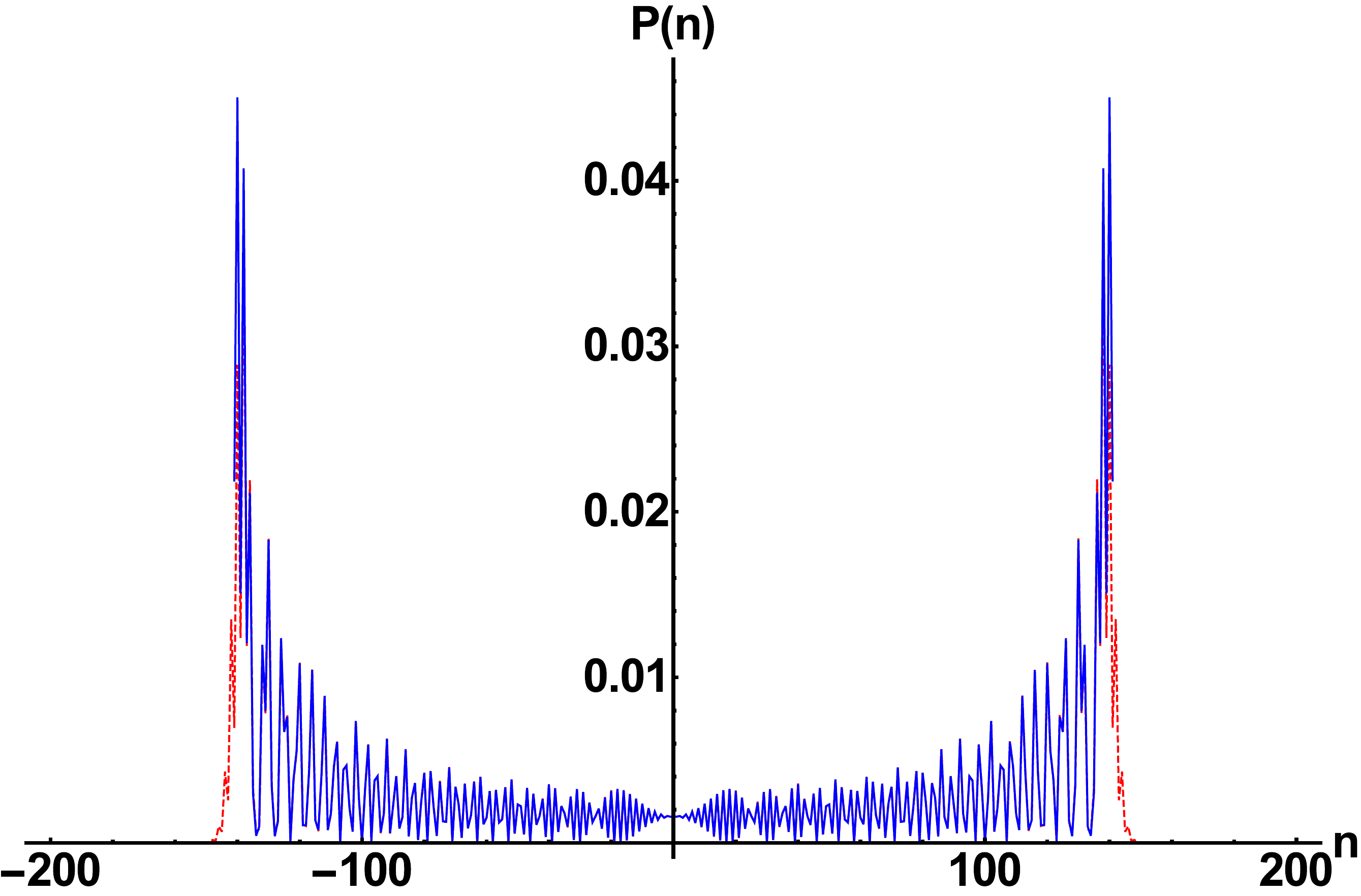}

\caption{(Color online) Comparison of the results from the stationary phase
method (solid blue line), with the exact evolution (dashed red line),
after $200$ time steps. The initial state is the localized initial
condition with $a=\frac{1}{\sqrt{2}}$, $b=\frac{i}{\sqrt{2}}$. }

\label{fig:probaprvsapr}
\end{figure}

We have represented in Fig. \ref{fig:probaprvsapr} the probability
distribution for the DQCA, as obtained from the stationary phase method,
compared with the exact evolution, starting from the localized initial
condition Eq. (\ref{eq:initialcodab}) with $a=\frac{1}{\sqrt{2}}$,
$b=\frac{i}{\sqrt{2}}$. The plot shows that this approximation works
very well, and accurately describes the oscillatory behavior of the
probability within the limits $|\alpha|\le\sqrt{1-\beta^{2}}$. A
detailed analysis shows that the differences in both curves are always
lower than $\sim5\%$ whenever $t\alpha\in\mathbb{Z}$ (i.e., for
points with support on the lattice).

The details of these oscillations are better seen for a simple case,
corresponding to the initial condition 
\begin{equation}
\psi(n,0)=\left(\begin{array}{c}
1\\
0
\end{array}\right)\delta_{n,0}.\label{eq:initial10}
\end{equation}
For this particular case, the probability distribution can be expressed,
after some algebra, as 
\begin{equation}
P(\alpha,t)=\frac{1}{\pi t\lambda^{''}(p_{s})}\left\{ (1+\alpha)^{2}[1+(-1)^{t+t\alpha}]\cos^{2}(t\phi(\alpha)+\pi/4)+(1-\alpha^{2})[1-(-1)^{t+t\alpha}]\sin^{2}(t\phi(\alpha)+\pi/4)\right\} .\label{eq:prob10DQCA}
\end{equation}
This result shows a clear difference with the corresponding result
for the QW (c.f. Eq (8) in \cite{Nayak2007}), where one obtains a
common factor $1+(-1)^{t+t\alpha}$, reflecting the parity properties
of the QW: for even (odd) $t$, the probability distribution vanishes
at odd (even) sites. Instead, Eq. (\ref{eq:prob10DQCA}) contains
a contribution of both even and odd sites at any time step $t$, as
detailed above.

We now return to Fig. \ref{fig:probaprvsapr} to make the following
observation. As already observed in Sect. II, the asymptotic form
of the DQCA looks very similar to the one obtained for the QW. One
may wonder about the mathematical reasons behind this resemblance.
The weak limit showed at the beginning of this Sect. can be used to
give an answer, as it manages to describe the overall shape of the
distribution. As can be seen from Eq. (\ref{eq:Exrwithp-1}), this
shape is governed, at late times, by the dispersion relation, which
can be made to coincide once the relevant parameters (the coin angle
for the QW, and parameter $\beta$ for the DQCA) are conveniently
mapped to each other using (\ref{eq:equaldisper}). Of course, one
has to remember that details in both distributions are different,
as discussed above. 

Similar conclusions can be reached within the stationary phase method.
The analysis that was used to obtain an approximate expression for
the functions $I_{i}(\alpha,t)$ starts with the expansion Eq. (\ref{eq:expansionphase}),
that only depends on the dispersion relation. Thus, these functions
are the same for the QW and the DQCA, once the mapping (\ref{eq:equaldisper})
is established. Now, which precise combinations of the real and imaginary
part of these functions one needs is dictated by the model (for the
DQCA, this combination is given by Eqs. (\ref{eq:psiRnt},\ref{eq:psiLnt}),
whereas for the QW one would need a different combination). Again,
this explains that we observe a similar shape in the probability distribution,
modulo the obtained differences.

\section{Entanglement}

A characteristic property of the QW is that entanglement between the
coin and spatial degrees of freedom is generated as a consequence
of the evolution \cite{1367-2630-7-1-156,Venegas-Andraca2005,Endrejat2005,PhysRevA.73.042302,PhysRevA.74.042304,Maloyer2007,PhysRevA.75.032351,PhysRevA.79.032312,Annabestani2010a}.
The amount of entanglement is usually quantified using the von Neumann
entropy of the reduced density matrix of the coin degrees of freedom,
after tracing out the spatial ones. More precisely, we define this
quantity, as a function of the time step $t$, by
\[
S(t)=-\Tr{\rho_{c}(t)\log_{2}\rho_{c}(t)},
\]
where $\rho_{c}(t)\equiv\sum_{n}\escp n{\psi(t)}\escp{\psi(t)}n$
is the reduced density matrix for the coin space, $Tr$ represents
the trace operation in this space, and $\log_{2}$ is the logarithm
in base $2$. 

\begin{figure}
\includegraphics[width=10cm]{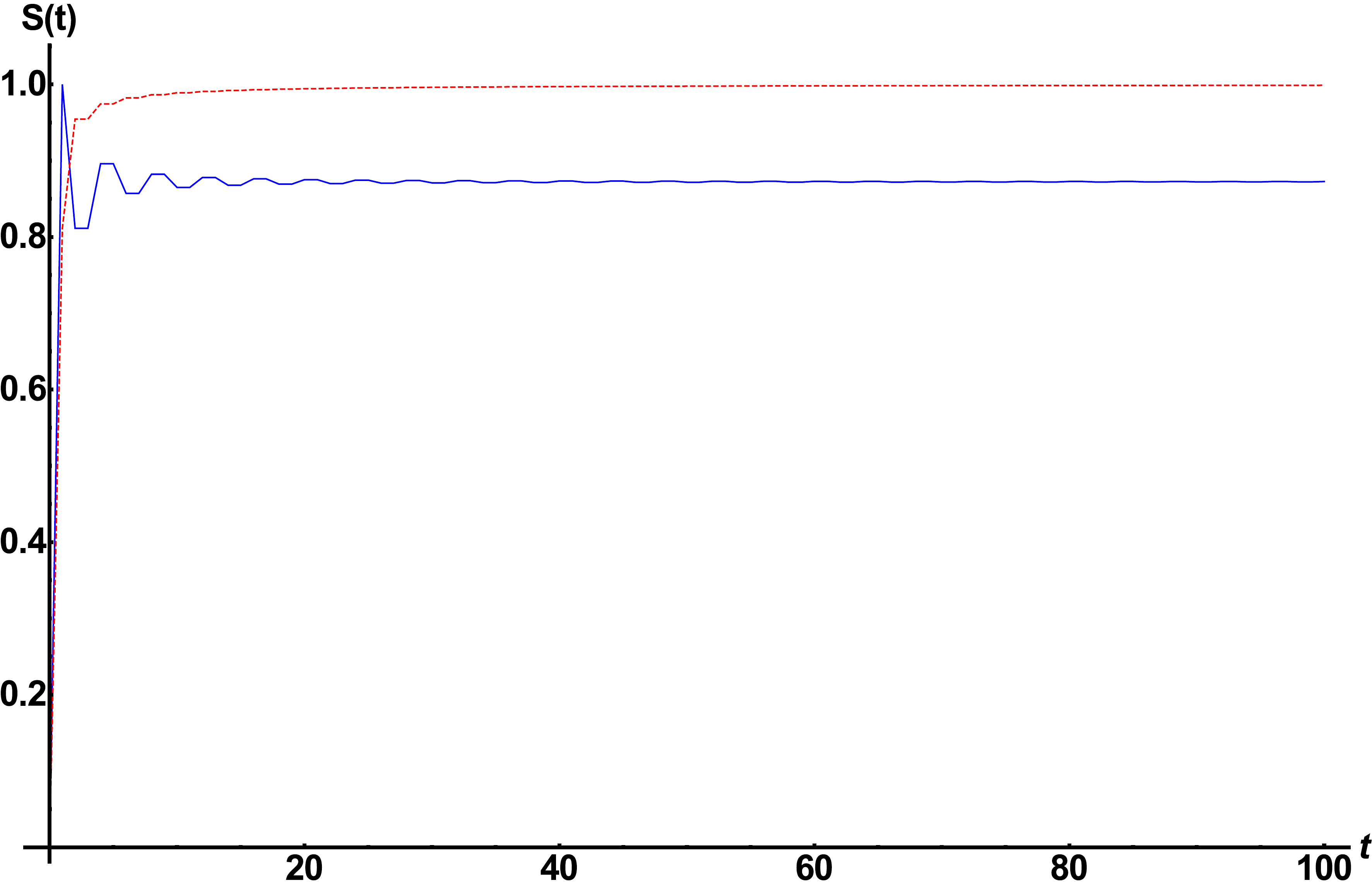}

\caption{(Color online) Entropy of entanglement as a function of the time step
$t$ for the QW (blue solid line) and for the DQCA (red dashed line).
The initial state in both cases is localized at the origin, Eq. (\ref{eq:initialdelta}). }

\label{fig:entroQWvsDQCA}
\end{figure}

As numerically obtained in \cite{1367-2630-7-1-156}, and proven later
in \cite{PhysRevA.73.042302}, for a Hadamard walk with localized
initial conditions the asymptotic entanglement is $S_{lim}\simeq0.8720$
for all initial coin states, although higher values can be reached
by starting from non-localized conditions (see also \cite{Hinarejos2013}).
An obvious question is whether the DQCA is also limited to this amount
of entanglement, when the evolution starts from the same state. Fig.
(\ref{fig:entroQWvsDQCA}) plots the entropy of entanglement $S(t)$
as a function of the time step for both the QW and the DQCA. We immediately
see that, for the QW, one approaches the predicted value $S_{lim}$.
Interestingly, the DQCA model overcomes this value, and reaches the
allowed maximum $S_{max}=1$ for a 2-dimensional system, thus indicating
that internal and motion degrees of freedom become maximally entangled.
Such large values of the entanglement are also reached for the one-dimensional
Dirac equation with narrow initial conditions, for some configurations
of the internal degrees of freedom, including the one used in Eq.
(\ref{eq:initialdelta}) \cite{Strauch2007}. As already discussed
in Sect. V, when considering highly localized states for the Dirac
equation, one has to be careful, since arbitrarily peaked states are
inconsistent with the one-particle approach \cite{Bracken2005}. One
can, however, consistently restrict to positive (or negative) energy
eigenvalues. For such states, the evolution of highly localized wave
packets gives rise to maximal entanglement.

We can get some insight into the above numerical results by obtaining
an analytical expression for the reduced density matrix $\rho_{c}(t)$
in the long term limit. This calculation is more conveniently done
in the quasimomentum space. We start from the initial state (\ref{eq:initialcodab}),
and make use of Eqs. (\ref{eq:PsiRp},\ref{eq:PsiLp}), which allows
us to obtain $\ket{\psi(p,t)}=(\psi_{R}(p,t),\psi_{L}(p,t))^{T}$.
Therefore we have
\begin{equation}
\rho_{c}(t)=\int_{-\pi}^{\pi}dp\ket{\psi(p,t)}\bra{\psi(p,t)}.
\end{equation}
After expansion, the matrix defined by $\ket{\psi(p,t)}\bra{\psi(p,t)}$contains
terms of the form $\sin\lambda(p)t\cos\lambda(p)t$, $\sin^{2}\lambda(p)t$,
and $\cos^{2}\lambda(p)t$. For large values of $t$, such terms become
highly oscillatory, while the rest of terms that depend on the variable
$p$ are smooth functions. Thus, we can replace the oscillatory contributions
by their averaged value: $\sin\lambda(p)t\cos\lambda(p)t\longrightarrow0$,
$\sin^{2}\lambda(p)t\longrightarrow1/2$, $\cos^{2}\lambda(p)t\longrightarrow1/2$.
The integral over the resulting expression becomes trivial, and we
finally obtain
\begin{equation}
\overset{\sim}{\rho_{c}}\equiv\lim_{t\rightarrow\infty}\rho_{c}(t)=\frac{1}{2}\left(\begin{array}{cc}
\beta\left|b\right|^{2}-(\beta-2)\left|a\right|^{2} & 2\beta Re(ab^{*})\\
2\beta Re(ab^{*}) & \beta\left|a\right|^{2}-(\beta-2)\left|b\right|^{2}
\end{array}\right).
\end{equation}
We can further represent the initial state on the Bloch sphere
\begin{equation}
\ket{\psi(p,0)}=\frac{1}{\sqrt{2\pi}}\binom{\cos\frac{\gamma}{2}}{e^{i\varphi}\sin\frac{\gamma}{2}},\label{eq:initialBloch}
\end{equation}
where $\gamma\in[0,\pi]$ and $\varphi\in[0,\pi]$. Then the above
result can be expressed as 
\begin{equation}
\overset{\sim}{\rho_{c}}=\frac{1}{2}\left(\begin{array}{cc}
1+(\beta-1)\cos\gamma & \beta\cos\varphi\sin\gamma\\
\beta\cos\varphi\sin\gamma & 1-(\beta-1)\cos\gamma
\end{array}\right).
\end{equation}
As observed from this expression, for the choice $\gamma=\pi/2$,
together with $\varphi=\pi/2,3\pi/2$ ones has 
\begin{equation}
\overset{\sim}{\rho_{c}}=\frac{1}{2}\left(\begin{array}{cc}
1 & 0\\
0 & 1
\end{array}\right),
\end{equation}
independently of the parameter $\beta$. For such values, then 
\begin{equation}
\lim_{t\rightarrow\infty}S(t)=1,
\end{equation}
which agrees with the result showed in Fig. (\ref{fig:entroQWvsDQCA}),
since the state Eq. (\ref{eq:initialdelta}) used for this calculation
can be described by the values $\gamma=\pi/2$, $\varphi=\pi/2$. 

The differences observed in the amount of entanglement generated within
the DQCA, as compared to the QW, may have important consequences.
The QW has been suggested as a possible device to generate entanglement
in quantum information processes \cite{Goyal2009}. On the other hand,
the coin can be regarded as a thermodynamic subsystem interacting
with the lattice. As such, it becomes an interesting scenario to investigate
the approach to thermodynamical equilibrium in quantum systems \cite{PhysRevA.85.012319}.
We have shown that the DQCA behaves differently to the QW, with a
dynamics that allows to reach the maximum allowed entanglement. Therefore,
it is possible that the transition towards equilibrium will show new
features. Among these features is the investigation of a non-Markovian
behavior previous to the asymptotic regime, as already observed for
the QW \cite{PhysRevA.89.052330}. All these perspectives clearly
deserve further investigation.

\section{Conclusions}

The connection of field theories on a lattice with simpler models
that can be used, in some limit, to simulate those theories, has proven
to be both a useful computational tool, and an avenue towards the
understanding of the underlying difficulties of the initial system.
In this work, we have investigated the time evolution of the Dirac
Quantum Cellular Automaton, initially proposed as a discretized version
that accounts for the motion of a relativistic spin 1/2 particle in
one dimension \cite{Bisio2015}, and compared its properties with
those of the Quantum Walk, an important primitive for quantum information.
We departed from the original motivation of the DQCA, and redefined
it as a discrete time quantum process taking place on an ordinary
lattice, analogously to the QW, that can in principle be implemented
using similar physical realizations, and allowing to compare both
processes on an equal footing. 

The probability distribution looks similar for both systems, with
some differences which arise from the fact that the DQCA includes
a term in the probability amplitude that forces the walker to stay
at the original position, at variance with the known properties of
the QW. In spite of these differences, both probability distributions
propagate in a similar manner, as clearly shown by the close resemblance
of the standard deviation in both cases. Given the analogy in the
propagation properties, one would expect similar capabilities in applications
to quantum algorithms. However, in order to reach the full potential
of the DQCA, one probably will need to generalize it to more general
graphs, as in the case of the discrete time QW, in order to look for
speedup in quantum search \cite{PhysRevA.67.052307,PhysRevA.78.012310},
element distinctness \cite{Ambainis1}, or even universal quantum
computation \cite{PhysRevA.81.042330}. 

We have given two analytic approximations to the probability distribution
at large time steps. The first one was obtained by calculating the
generalized momentum of the position operator. In this way, one obtains
a simple result that only describes the general shape of the distribution,
although it suffices to take account for the observed ballistic evolution
of the standard deviation. On the other hand, the stationary phase
method provides a good approximation, and clearly shows the effect
of the above mentioned ``probability to stay'' term in the DQCA.

The analysis of the entanglement between the internal and spatial
degrees of freedom reveals that, for some initial choices of the coin
state localized at one point of the lattice, the DQCA approaches a
maximally entangled state . This result clearly overcomes the known
limiting values of the QW for similar initial conditions. Maximally
entangled states also appear for some narrow solutions of the Dirac
equation in the continuum so that, in this respect, the DQCA looks
closer to a Dirac particle than the QW.

To summarize, the DQCA can be regarded as an alternative to the QW,
which shares many properties with it, while possessing some new distinctive
features. At the same time, given its original motivation, it can
easily serve as a model to illustrate many properties of the Dirac
equation, such as the Zitterbewegung and scattering from a potential
\cite{Bisio2013}. Of course, an important point is the possibility
of experimentally realizing the DQCA. In this respect, one might look
for setups similar to the ones used to realize the QW, or perhaps
make use of some recent ideas on quantum dots to implement quantum
cellular automata \cite{Imre13012006,rohtua,Tougaw1994}. 
\begin{acknowledgments}
This work has been supported by the Spanish Ministerio de Educación
e Innovación, MICIN-FEDER project FPA2011-23897 and FPA2014-54459-P,
SEV-2014-0398 and ``Generalitat Valenciana'' grant GVPROMETEOII2014-087.
The author acknowledges illuminating discussions with E. Roldán and
G. de Valcárcel.
\end{acknowledgments}

\end{document}